\documentstyle[NATO,epsf]{CRCKAPB}

\begin{opening}
\title{Stellar wind signatures in sdB stars?}
\author{U. Heber}
\institute{Dr. Remeis-Sternwarte, Astronomisches Institut, Universit\"at 
Erlangen-N\"urnberg, Bamberg, Germany}
\author{P.F.L. Maxted}
\institute{School of Chemistry\,\&\,Physics, Keele University,\,GB}
\author{T.R. Marsh, C. Knigge}
\institute{University of Southampton, Dept. of Physics\,\&\,Astronomy,\,GB}
\author{J. E. Drew}
\institute{Imperial College, London, GB}

\end{opening}

\begin{document}

% The \begin{document} command comes after the \end{opening}
% command.
Subdwarf B (sdB) stars form the blue end of the horizonal branch (EHB). Their 
peculiar atmospheric abundance patterns are due to diffusion processes.
However, diffusion models fail to explain these anomalies quantitatively.
From a NLTE model atmosphere analysis of 40 sdB stars,   
Heber et al. \cite{heber03} found that the four most luminous stars in 
their sample have already evolved off the EHB (post-EHB) and display 
anomalous H$\alpha$ and HeI\,6678\AA\ line profiles, i.e, the lines are too
broad and shallow and may even show some emission. For illustration we 
compare in Fig. 1
the spectrum of a post-EHB stars (PG~1000$+$408) to that of a 
typical EHB stars (PG~1110$+$294). Note that 
synthetic line profiles displayed are not fits to the observed line profiles 
but are calculated from models with atmospheric parameters that have been 
predetermined by matching their blue spectra 
\cite{maxted01}. The observed H$\alpha$ and HeI\,6678\AA\ line profiles of 
all EHB stars, however, are very well reproduced by the 
synthetic spectra as is demonstrated in Fig.~1 for PG~1110$+$294.

What is the reason for the line profile anomaly for the more luminous sdB 
stars?
The spectral analysis was carried out using static, hydrogen and helium 
line blanketed model atmospheres.
There are two assumptions in the models that might not hold for the 
four luminous post-EHB stars. 
The influence of metal lines on the atmospheric structure has been neglected 
and hence metal line blanketing could alter the synthetic spectrum 
more severely for the post-EHB stars than for the EHB stars. 
However, metal lines are very weak or absent in the available medium 
resolution optical spectra of the post-EHB
stars and are even weaker than in a typical EHB star.   
Alternatively, assuming hydrostatic equillibrium might be 
incorrect.
Mass loss rates of radiatively driven winds increase non-linearly with 
increasing stellar 
luminosity. Therefore we expect stellar winds to effect the most luminous 
stars in our sample the most just as it is observed. 
Hence we regard the observed spectral anomalies
as signatures of a stellar wind, the first such detection in
this class of star (if confirmed).  
High-quality UV spectra are needed to derive more stringent limits on the 
mass loss rates as well as on the metal abundances of the programme stars.
Recent model calculations for globular cluster HB stars of late B-type 
\cite{vink02}, 
which accounted for stellar winds, provided encouraging progress
towards a solution of
the zoo of problems associated with globular clusters' blue HBs. 
However those stars are considerably cooler than our programme stars.
Similar calculations of synthetic spectra for the latter are urgently needed 
to validate the stellar wind hypothesis.   
% =================================================================================
\begin{figure}
\begin{center}
\vspace*{6.0cm}
\includegraphics{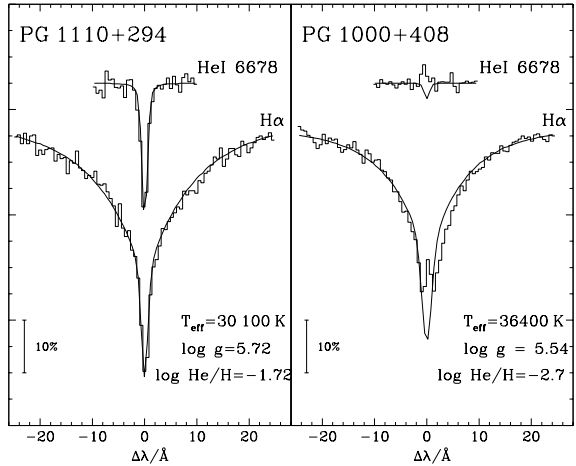}
\caption{Comparison of synthetic line profiles of H$\alpha$ and 
He I, 6678\AA\ to the observations of a low luminosity sdB star on the EHB 
(left) and a more luminous post-EHB star (right).}
\end{center}
\end{figure}
% =================================================================================

% ==================================================================================

% ====================================================================

\begin{thebibliography}{}
\bibitem[2003]{heber03}
Heber U., Maxted P.F.L., Marsh T.R., Knigge C., Drew J.E. 2003,  
Proc. of a {\it Workshop on Stellar Atmosphere Modelling}, 
held 8--12 April 2002 in T\"ubingen, Germany,  
eds. I. Hubeny, D. Mihalas $\&$ K. Werner, PASPC, in press
%\bibitem{hua00}
%Hui-Bon-Hoa A., LeBlanc, F. Hauschildt P. H. 2000, ApJ 553, L43 
\bibitem[Maxted et al. 2001]{maxted01}
Maxted P.F.L., Heber U., Marsh T.R., North R.C. 2001, MNRAS 326,1391
\bibitem[Vink \& Cassisi 2002]{vink02}
Vink J., Cassisi S. 2002, A\&A in press (astro-ph/0207037)
\end{thebibliography}
\end{document}